\documentclass[journal=nalefd,manuscript=letter,layout=traditional]{achemso}
\usepackage[version=3]{mhchem} % Formula subscripts using \ce{}
\usepackage[T1]{fontenc}       % Use modern font encodings
\usepackage{amsmath}
\usepackage{amssymb}
\usepackage{bm}
\usepackage{braket}
\usepackage{graphicx}
\usepackage{comment}
\usepackage{soul}
\usepackage{color}

\author{Yi Yang}
\affiliation{Research Laboratory of Electronics, Massachusetts Institute of Technology, Cambridge, Massachusetts 02139, USA}
\email{yiy@mit.edu}
\author{Bo Zhen}
\affiliation{Research Laboratory of Electronics, Massachusetts Institute of Technology, Cambridge, Massachusetts 02139, USA}
\alsoaffiliation{Physics Department and Solid State Institute, Technion, Haifa 320000, Israel}
\author{Chia Wei Hsu}
\affiliation{Department of Applied Physics, Yale University, New Haven, CT 06520, USA}
\author{Owen D. Miller}
\affiliation{Department of Mathematics, Massachusetts Institute of Technology, Cambridge, MA 02139, USA}
\author{John D. Joannopoulos}
\affiliation{Research Laboratory of Electronics, Massachusetts Institute of Technology, Cambridge, Massachusetts 02139, USA}
\author{Marin Solja\v{c}i\'{c}}
\affiliation{Research Laboratory of Electronics, Massachusetts Institute of Technology, Cambridge, Massachusetts 02139, USA}

\title{Optically Thin Metallic Films for High-radiative-efficiency Plasmonics}


\begin{document}

%%%%%%%%%%%%%%%%%%%%%%%%%%%%%%%%%%%%%%%%%%%%%%%%%%%%%%%%%%%%%%%%%%%%%%
%%% The "tocentry" environment can be used to create an entry for the
%%% graphical table of contents. It is given here as some journals
%%% require that it is printed as part of the abstract page. It will
%%% be automatically moved as appropriate.
%%%%%%%%%%%%%%%%%%%%%%%%%%%%%%%%%%%%%%%%%%%%%%%%%%%%%%%%%%%%%%%%%%%%%%
%\begin{tocentry}
%
%Some journals require a graphical entry for the Table of Contents.
%This should be laid out ``print ready'' so that the sizing of the
%Inside the \texttt{tocentry} environment, the font used is Helvetica
%8\,pt, as required by \emph{Journal of the American Chemical
%The surrounding frame is 9\,cm by 3.5\,cm, which is the maximum
%permitted for  \emph{Journal of the American Chemical Society}
%graphical table of content entries. The box will not resize if the
%content is too big: instead it will overflow the edge of the box.
%
%This box and the associated title will always be printed on a
%separate page at the end of the document.
%
%\end{tocentry}

\newpage
\begin{abstract}
Plasmonics enables deep-subwavelength concentration of light and has become important for fundamental studies as well as real-life applications.
Two major existing platforms of plasmonics are metallic nanoparticles and metallic films.
Metallic nanoparticles allow efficient coupling to far field radiation, yet their synthesis typically leads to poor material quality.
Metallic films offer substantially higher quality materials, but their coupling to radiation is typically jeopardized due to the large momentum mismatch with free space.
Here, we propose and theoretically investigate optically thin metallic films as an ideal platform for high-radiative-efficiency plasmonics.
For far-field scattering, adding a thin high-quality metallic substrate enables a higher quality factor while maintaining the localization and tunability that the nanoparticle provides. For near-field spontaneous emission, a thin metallic substrate, of high quality or not, greatly improves the field overlap between the emitter environment and propagating surface plasmons, enabling high-Purcell (total enhancement > $10^4$), high-quantum-yield (> 50\%) spontaneous emission, even as the gap size vanishes (3$\sim$5 nm).
The enhancement has almost spatially independent efficiency and does not suffer from quenching effects that commonly exist in previous structures.
\end{abstract}

\newpage

Ohmic loss in metals is the most critical restriction for plasmonics \cite{khurgin2015deal}. The restriction can be characterized by the radiative efficiency $\eta$, defined as the ratio between the radiative decay rate and the total decay rate, i.e., $\eta=\gamma_{\text{rad}}/\gamma_{\text{tot}}$. Two major existing platforms of plasmonics are metallic nanoparticles\cite{sobhani2015pronounced,liu2012broadband,chang2011low,peer2007nanocarriers,ota2013lipid,maier2003local} and metallic films \cite{wu2014intrinsic,mcpeak2015plasmonic,babar2015optical,miller2014effectiveness}; they both face their own restrictions for achieving a high $\eta$. A major problem regarding nanoparticles is their poor material qualities due to the amorphous structures that arise from the colloidal synthesis processes. In comparison, single- or poly-crystalline metallic films fabricated via temperature-controlled sputtering or epitaxial growth can achieve much higher material qualities and much lower material losses, but their coupling to radiation is typically jeopardized due to the large momentum mismatch with free space. When the two platforms are combined, the radiation of nanoparticles is also at risk of being quenched by a bulk nearby metallic film. These restrictions lead to compromises between $\eta$ and other mode properties, such as quality factor ($Q$) and mode volume \cite{koenderink2010use,sauvan2013theory,kristensen2013modes} ($V$).

For plasmonic light scattering, it is often desirable to achieve high radiative efficiencies and high $Q$ simultaneously.
In biomedical sensing \cite{lee2005dependence,el2005surface,anker2008biosensing,saha2012gold}, for example, a high Q is required for high spectral resolution, while a high radiative efficiency (stronger scattering) is needed for high signal-to-noise ratio (SNR).
Meanwhile, transparent displays\cite{hsu2014transparent,hsu2015optimization,saito2015transparent} based on resonant scattering demand high $Q$ for high transparency and high radiative efficiencies for high brightness.
However, it is very challenging to achieve both goals at the same time. First, $Q$, $\sigma_{\text{ext}}$, and $\sigma_{\text{sca}}$ are all bounded from above as functions of the permittivities of materials \cite{PhysRevLett.97.206806,PhysRevLett.110.183901,PhysRevLett.112.123903,sigmaLimitArxiv}, primarily due to the intrinsic material loss. Second, there exists a fundamental physical contradiction between the two requirements: higher radiative efficiencies require higher radiative decay rates, which necessarily reduce the total quality factors.

For plasmon-enhanced emission\cite{anger2006enhancement,kuhn2006enhancement,russell2012large,rose2014control,akselrod2014probing,eggleston2015optical,pelton2015modified}, another trade-off exists between achieving high quantum yield (QY) and large Purcell \cite{purcell1946spontaneous} factors, even though both are typically desired.
The key to achieving high spontaneous emission enhancement over a broad band \cite{pelton2015modified,vesseur2010broadband} using plasmonics is to achieve small $V$s.
However, as $V$ decreases, absorptive decay rates (proportional to $V$ \cite{bohren2008absorption}) dominate over radiative decay rates (proportional to $V^{2}$ \cite{bohren2008absorption}), triggering a drastic drop in QY \cite{faggiani2015quenching,eggleston2015optical,novotny2012principles}.
Recently, much effort has been made to enhance spontaneous emission using gap plasmons \cite{esteban2010optical,moreau2012controlled,russell2012large,belacel2013controlling,rose2014control,akselrod2014probing,faggiani2015quenching}, created via the confinement of light within the dielectric gap between nanoparticles and an optically thick metallic substrate.
Compared with other types of resonances, the gap plasmon resonance achieves high total enhancement \cite{akselrod2014probing} as it offers more reliable control of the dielectric gap thinness.
However, these gap plasmon resonances cannot circumvent the tradeoff between QY and $V$. For example, when the gap size is reduced to 5 nm or smaller for a nanocube, despite a higher total decay rate, the efficiency (defined as the sum of photon and plasmon radiative efficiency \cite{akselrod2014probing,rose2014control,faggiani2015quenching}) drops below $\sim$20\% \cite{akselrod2014probing,faggiani2015quenching}.
Moreover, the efficiency is strongly dependent on the location of emitters. QY reaches maximum if the emitter is placed at the center of the gap but decreases immensely when the emitter is in the proximity of the metal \cite{akselrod2014probing}.

Here, we propose and theoretically demonstrate that an optically thin metallic film makes an ideal platform for high-radiative-efficiency plasmonics via two examples: high-$Q$ scattering and enhanced emission.
For scattering, a high-quality thin metallic film facilitates a high-$Q$, high-radiative-efficiency Mie plasmon resonance, whose $Q$ exceeds the quasistatic $Q$ of the nanoparticle material.
For enhanced emission, gap plasmons can still be well supported and are better mode overlapped with external radiation using an optically-thin metallic substrate.
A high-Purcell (total enhancement > $10^4$), spatially-independent-efficiency (>50\%) spontaneous emission enhancement can be achieved with vanishing gap size (3$\sim$5 nm), even if the substrate has the same material properties as the nanoparticles.
Our platform can also be extended to other applications (for example, nonlinear frequency generation and multiplexing), because of the enhanced efficiencies of high-order plasmonic modes. Moreover, the ratio between photon and plasmon radiation can be easily tailored by altering the shape of the nanoparticles, making this platform versatile for both fluorescence \cite{akselrod2014probing,rose2014control,eggleston2015optical} and plasmon circuits \cite{koenderink2009plasmon,gonzalez2011entanglement,gan2012proposal,kumar2014generation}.

Below we show that in plasmonic optical scattering, the quasistatic $Q$ of a deep subwavelength nanoparticles can be exceeded with the help of an optically thin high-quality metal film, while maintaining considerably high radiative efficiencies $\eta$, which is also known as the scattering quantum yield \cite{lee2005dependence} or the albedo \cite{newton2013scattering} in scattering problems.
For a subwavelength scattering process, based on temporal coupled-mode theory\cite{PhysRevA.75.053801,PhysRevA.85.043828}, the radiative efficiency $\eta$ and the total quality factor $Q_{\text{tot}}$ for a single resonance are given by
\begin{eqnarray}
\eta&\equiv&\frac{\gamma_{\text {rad}}}{\gamma_{\text {tot}}}=\frac{\sigma_{\text {sca}}}{\sigma_{\text {ext}}}  \bigg|_{\textrm{on resonance}},\label{eff}\\
Q_{\text {tot}} &=& \omega_0/2{\gamma_{\text {tot}}}, \label{eqQ}
\end{eqnarray}
where $\omega_0$ is the resonant frequency, $\gamma_{\text{tot}}=\gamma_{\text{rad}}+\gamma_{\text{abs}}$ is the total decay rate, and $\sigma_{\text {ext}}=\sigma_{\text {sca}}+\sigma_{\text {abs}}$ is the extinction cross-section. As $\gamma_{\text {abs}}$ is mostly dictated by material absorption \cite{PhysRevLett.97.206806,PhysRevLett.110.183901}, to get a high $\eta$, one has to increase $\gamma_{\text {rad}}$. This in turn spoils the quality factor (Eq. \ref{eqQ}), which reveals the trade-off between $\eta$ and $Q_{\text {tot}}$, as we described previously.
Because simultaneously achieving a high $Q$ and a high $\eta$ is important for many applications, like biomedical sensing\cite{lee2005dependence,el2005surface,anker2008biosensing,saha2012gold} and transparent displays\cite{hsu2014transparent,hsu2015optimization,saito2015transparent}, we define the figure of merit (FOM) for scattering as
\begin{eqnarray}
\text {FOM}_{\text {sca}} =\frac{Q_{\text {tot}}}{1-\eta}. \label{fom1}
\end{eqnarray}
It follows that this FOM reduces to the quasistatic quality factor $Q_{\text {qs}}$ \cite{PhysRevLett.97.206806}
\begin{eqnarray}
\text {FOM}_{\text {sca}}=\omega_0\slash{2\gamma_{\text {abs}}}=Q_{\text{abs}}\simeq Q_{\text {qs}}=\frac{\omega\frac{d\epsilon'}{d\omega}}{2\epsilon''}, \label{fom2}
\end{eqnarray}
which only depends on the material property of the nanoparticle. Here, $\epsilon'$ and $\epsilon''$ are real and imaginary parts of the complex permittivity. For subwavelength metallic nanoparticles (dimension $\ll \lambda$), their plasmon properties are typically dominated by quasistatic considerations\cite{PhysRevLett.97.206806} and thus the approximation $Q_{\text{abs}}\simeq Q_{\text{qs}}$ holds, which also indicates that the material loss inside the metallic nanoparticle cannot be further reduced. Therefore, our strategy is to squeeze parts of the resonant mode into a high-quality metallic film\cite{wu2014intrinsic,mcpeak2015plasmonic} with much lower loss, while maintaining efficient radiation rates.

\begin{figure}[hbtp]
  \centering
  \includegraphics[width=6 in]{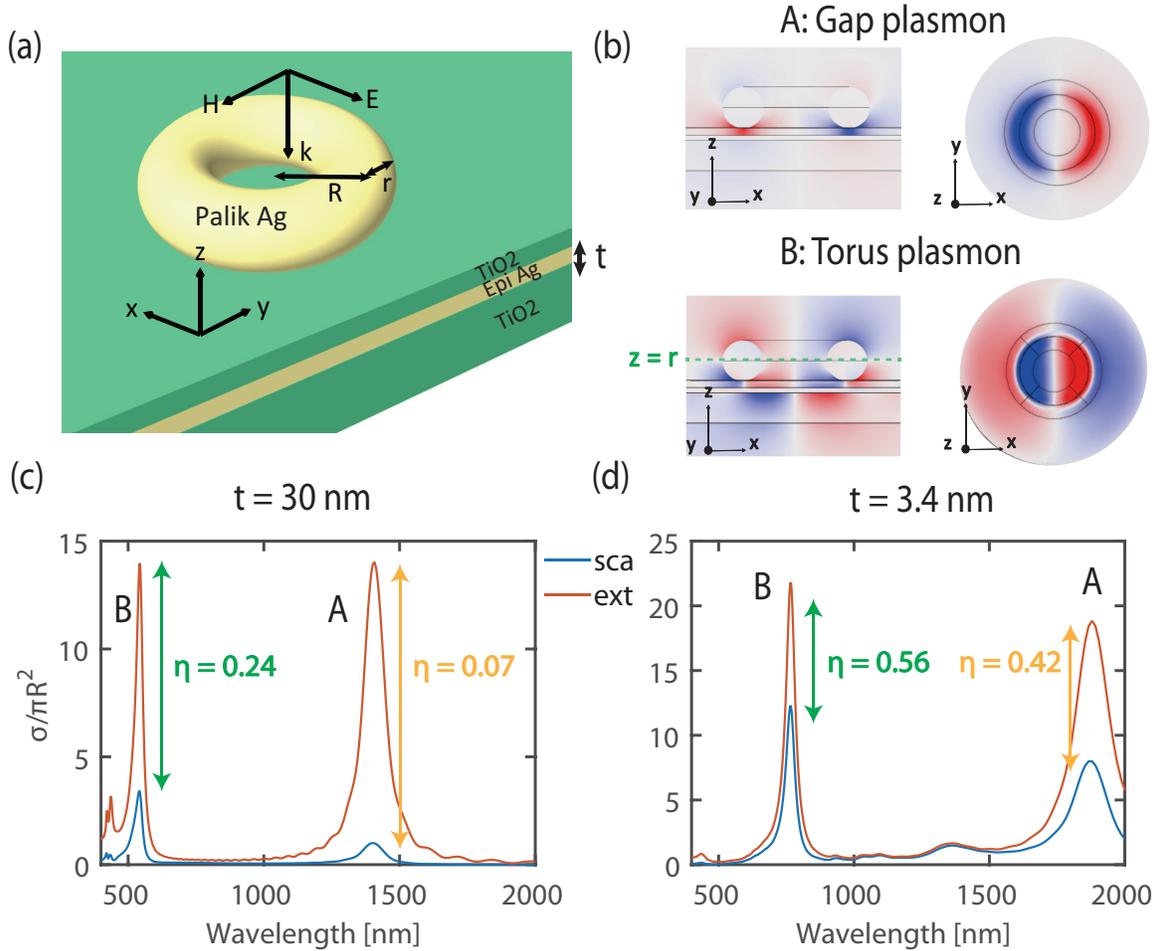}
  \caption{(a) Structure: a torus sitting on top of a metallic multifilm. The major and minor (cross-section) radii are denoted by $R=36$ nm and $r=14$ nm, respectively. The thicknesses of the upper and lower amorphous TiO$_2$ layers are fixed at 5 nm and 20 nm respectively. The thickness of the middle epitaxial silver layer is denoted by $t$. (b) $E_z$ profiles of two eigenmodes when $t=3.4$ nm in $x-z$ (left) and $x-y$ (right) planes. Upper: gap plasmon resonance; Lower: torus (Mie) plasmon resonance. Scattering and extinction cross-sections of the torus on a (c) thick metal film ($t$ = 30 nm) and (d) thin metal film ($t$ = 3.4 nm), respectively. The radiative efficiency $\eta$ increases significantly when metal thickness is reduced.
}
  \label{sca1}
\end{figure}

As an example, we investigate a silver torus \cite{mary2005localized,dutta2008plasmonic,teperik2011numerical,rakovich2015plasmonic} scatterer, sitting on top of a TiO$_2$-Ag-TiO$_2$ multifilm, whose structural geometry is shown in Fig.~\ref{sca1}(a). The permittivities of the silver film and the torus are obtained from Wu\cite{wu2014intrinsic} and Palik\cite{palik1998handbook}, respectively; the former has substantially lower loss since it is assumed to be made epitaxially. The permittivity of amorphous TiO$_2$ (refractive index $\sim$ 2.5 in the visible and near-infrared spectra) is from Kim \cite{kim1996simultaneous}. The material absorption in TiO$_2$ is negligible compared with the absorption in silver, as $\text {Im}(\epsilon_{\text {TiO$_2$}})$ is several orders of magnitude lower than that of $\text {Im}(\epsilon_{\text {Ag}})$ within the wavelength range of interest. Thus the absorption in TiO$_2$ is not considered in the calculation.
The ambient index of refraction is $1.38$ (near the refractive index of water, tissue fluids, and various polymers). If the structure is probed with normally incident plane waves, only the $m=1$ ($m$ is the azimuthal index of the modes since the structure is axially-symmetric) modes of the structure can be excited \cite{bohren2008absorption}. Fig.~\ref{sca1}(b) shows the mode profiles of the two $m=1$ resonances in this structure. Resonance A is a gap plasmon resonance \cite{moreau2012controlled} whose field is mostly confined in the upper TiO$_2$ layer. Resonance B corresponds to the torus (Mie) plasmon resonance \cite{yamamoto2010gap}, given that it maintains a nodal line [green dashed line in Fig.~\ref{sca1}(b)] along $z=r$ ($r$ is the minor radius of the torus), which is a feature of the torus resonance in free space \cite{mary2005localized,dutta2008plasmonic,teperik2011numerical,rakovich2015plasmonic}.
Fig.~\ref{sca1}(c) and (d) compare $\sigma_{\text{sca}}$ and $\sigma_{\text{ext}}$ of the torus when the silver layer in the multifilm is optically thick ($t$ = 30 nm) or thin ($t$ = 3.4 nm). For both resonances, the radiative efficiency in the thin-film case is much higher than that in the thick-film case.
Moreover, when the torus moves away from the multifilm, the response of resonances is very different for the thin film case from that for the thick film, as shown in Fig.~S1.
We now focus on the Mie resonance B for high-$Q$ scattering as most of its entire radiation (photon and plasmon combined) goes into the far field (photon).
We will return to the gap plasmon resonance A later for enhanced emission applications.

\begin{figure}[hbtp]
  \centering
  \includegraphics[width=3 in]{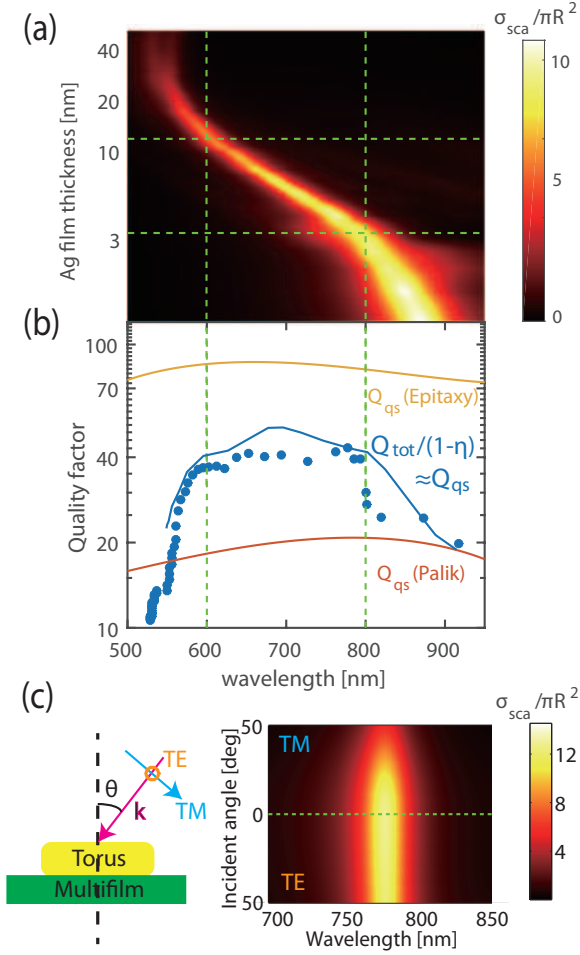}
  \caption{ (a) The scattering cross section $\sigma_{sca}$ of torus plasmon resonance decreases as the silver film thickness $t$ increases. (b) $\text{FOM}_{\text {sca}}=Q_{\text {tot}}\slash(1-\eta)\simeq Q_{\text{qs}}$ shows that our structure can exceed the quasistatic limits for the Palik silver used in the nanoparticle. When the silver film is optically thin ($t=3\sim10$ nm), a plateau of FOM$_{\text{sca}}\sim$40 exceeding quasistatic limit of the Palik silver is achieved for resonant wavelengths at 600$\sim$800 nm, as denoted by the dashed green lines. The blue dots are calculated via Eq. \ref{fom1} from the time-domain scattering simulation. The blue line is calculated via Eq. \ref{Qmodel} from the frequency-domain eigenmode simulation. (c) Angular dependence of the scattering cross section of the torus plasmon resonance with $t=3.4$ nm under the excitation of TE and TM polarizations.}
  \label{sca2}
\end{figure}

By changing $t$ from 0 nm to 50 nm while keeping other parameters unchanged ($t=0$ nm corresponds to a single 25-nm TiO$_2$ layer), we are able to track the torus plasmon resonance B and evaluate its FOM$_{\text {sca}}$, as shown in Fig.~\ref{sca2}. As $t$ increases, the resonance blueshifts, along with a reduced linewidth [Fig.~\ref{sca2}(a)].
In Fig.~\ref{sca2}(b), we compare the FOM$_{\text {sca}}$ in our structure to the quasistatic limit $Q_{\text {qs}}$ for different materials in the system: the Palik silver\cite{palik1998handbook} that is used for the torus and the epitaxial silver that is used for the substrate\cite{wu2014intrinsic} (FOM$_{\text {sca}}$ and $Q_{\text {qs}}$ are directly comparable, see Eqs. \ref{fom1} and \ref{fom2}).
There exists a plateau of higher FOM$_{\text {sca}}$ at $t=3\sim10$ nm.
At these thicknesses, the multifilm still has very high transmission { > 80\% }(Fig.~S2).
The FOM$_{\text{sca}}$ of the torus plasmon resonance exceeds and becomes twice as high as the $Q_{\text {qs}}$ of the torus material (Palik \cite{palik1998handbook}).
When the silver layer is either too thin (< 3 nm) or too thick (> 20 nm), the FOM$_{\text{sca}}$ drops considerably and FOM$_{\text{sca}}\lesssim Q_{\text {qs}}(\text{Palik})$, the quasistatic quality factor of the torus material. 
Fig.~\ref{sca2}(c) shows that the high FOM$_{\text {sca}}$ can be maintained for both polarizations over a wide range of incident angles.

The increased quality factor is the result of effective mode squeezing that only occurs in thin silver films -- an effect we qualitatively demonstrate in Fig.~S3 of the Supporting Information. The mode squeezing mechanism can be quantitatively demonstrated by calculating the energy density integral of the eigenmode. The energy density $u$ in lossy media is generally defined as $u = \epsilon_0\left(\epsilon'+2\omega\epsilon''/\gamma\right)|E|^2/2$ \cite{ruppin2002electromagnetic}, where $\epsilon'$ and $\epsilon''$ are real and imaginary parts of permittivity respectively, and $\gamma$ is the damping of the metal. We adopt $\gamma = 1.4\times10^{14}$ rad/s for the Palik silver and $\gamma = 3.14\times10^{13}$ rad/s for the epitaxial silver to best match the tabulated data. Since the metallic objects (Palik silver torus and epitaxial silver film) dominate the absorption loss in this system, we define the energy concentration coefficients in the torus and the film as
\begin{eqnarray}
c_{\text {torus}} = \frac{\int_\text{torus}u dV}{\int_\text{torus}u dV + \int_\text{film}u dV},\\
c_{\text {film}} = \frac{\int_\text{film}u dV}{\int_\text{torus}u dV + \int_\text{film}u dV}.
\end{eqnarray}
Thus, the $Q_{\text {qs}}$ of the system can be estimated as
\begin{eqnarray}
\frac{1}{Q_{\text{qs}}} = \frac{c_{\text {torus}}}{Q_{\text{qs}}(\text{Palik})} + \frac{c_{\text {film}}}{Q_{\text{qs}}(\text{Epitaxy})}. \label{Qmodel}
\end{eqnarray}
As shown in Fig.~\ref{sca2}(b), the $Q_{\text {qs}}$ of the system, calculated from the scattering (blue dots) and eigenmode (blue curve) simulations respectively, match each other well. Our calculation shows the high energy concentration in the film only happens when the film is optically thin (see Fig.~S4). Near the maximum of the $Q_\text{qs}$ (wavelength $\sim$700 nm, silver film thickness $\sim$7 nm), the energy concentrated in the film is three times higher than that in the torus ($c_{\text {film}} \sim 3c_{\text {torus}}$).
We also note that the $Q_{\text {qs}}$ curves of the two materials are quite flat within the wavelength of interest. Thus, it is the effective mode squeezing into a high-quality film, rather than the dispersion of an individual material, that contributes to the improved quality factor of the system.

The aforementioned enhanced $Q$ is different from the linewidth narrowing that is based on the interference between multiple resonances \cite{sobhani2015pronounced}. For coupled resonances, as the trace of the full Hamiltonian is conserved, the linewidth reduction of one resonance necessarily implies the broadening of the others'.
This coupling also typically renders the spectrum Fano-like with dark states in the middle of the spectrum \cite{hsu2014theoretical}.
In contrast, here the linewidth reduction is realized via effectively squeezing a single Mie plasmon mode into an optically-thin metallic film. Scattering spectrum is kept single-Lorentzian, which is favorable for many applications \cite{el2005surface,anker2008biosensing,saha2012gold,hsu2014transparent,saito2015transparent} as it maintains a high resolution and SNR. Moreover, as the resonance for scattering uses the Mie plasmon and the ambient environment is the perturbed free space, most of the reradiated energy goes into the far field with weak plasmon excitation (see supporting information).
We also note that optically thin metallic films are not restricted to high-$Q$ applications shown above. Applications based on broadband strong scattering (like solar cells requiring longer optical path) can also be implemented on this platform, utilizing its high radiative efficiency.

Antennas work equally well as receivers and as transmitters;
in the context of nanoparticles, the radiative efficiency $\eta$ is equally important, whether nanoparticles are used to scatter light from the far field or serve as external cavities to enhance spontaneous emission in the near field.
The quantum yield (QY) of an emitter (whose total decay rate is $\Gamma_0$ in free space) enhanced by a plasmonic nanoparticle can be approximated as\cite{pelton2015modified}
$
\text{QY}\simeq\eta\Gamma_{\text g}\slash\Gamma_{\text {tot}}
$
under the assumption that the decay rate is dominated by the plasmonic resonance (note we use $\Gamma$ and $\gamma$ to denote the emission and scattering processes respectively). Here, $\Gamma_{\text{tot}}=\Gamma_{\text{g}} + \Gamma_{0}' + \Gamma_{\text{nr}}^{\text{em}} + \Gamma_{\text{q}}$, $\Gamma_0'$ is the radiative decay rate of the emitter not coupled to the cavity, $\Gamma_{\text{g}}\simeq\Gamma_{\text{rad}}+\Gamma_{\text{abs}}$ is the modified emission rate in the presence of the cavity, $\Gamma_{\text{rad}}$ and $\Gamma_{\text{abs}}$ are radiative and absorptive decay rates of the cavity respectively,
$\Gamma_{\text{nr}}^{\text{em}}$ is the intrinsic nonradiative decay rate of the emitter, and $\Gamma_{\text{q}}$ is the quenching rate that refers to the loss induced by the direct heating of the metal from the emitter without coupling to optical resonances.
In most cases, $\Gamma_{\text {g}}$ is dominant over all other components of $\Gamma_{\text{tot}}$ and $\Gamma_{\text{rad}}$ is much larger than $\Gamma_0'$; therefore, we can further approximate QY as the radiative efficiency of the nanoparticle, i.e., $\text{QY}\simeq \eta$.
For enhanced emission, it is often desired to simultaneously achieve high quantum yield and high decay rates, so we define the FOM for enhanced emission as
\begin{eqnarray}
\text{FOM}_{\text{emit}}= \eta\cdot F_p \propto \eta/V, \label{eqemit}
\end{eqnarray}
where $F_p=\Gamma_{\text{tot}}/\Gamma_0$ is the Purcell factor \cite{purcell1946spontaneous} and $V$ is the mode volume \cite{koenderink2010use,sauvan2013theory,kristensen2013modes}. Note that $Q$ does not show explicitly in Eq.~\ref{eqemit} as the broadband plasmonic enhancement relies on $V$ much more than on $Q$. It follows that FOM$_{\text{emit}}$ reduces to the radiative enhancement $\Gamma_{\text{rad}}/\Gamma_0$.

\begin{figure}[hbtp]
  \centering
  \includegraphics[width=6 in]{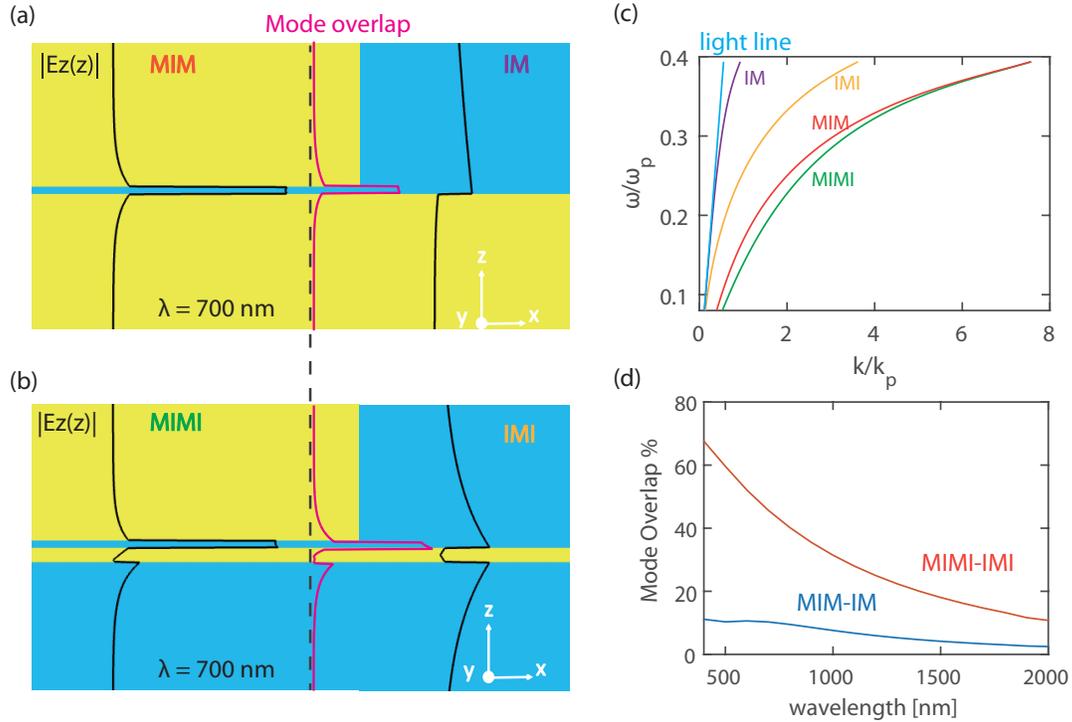}
  \caption{Mode-overlap analysis showing the advantage of using optically thin substrates for gap plasmon emission enhancement. Improved mode matching of surface plasmon polaritons (SPPs) comparing (a) the metal-insulator-metal and insulator-metal (MIM-IM) interface with a 12\% overlap to (b) the metal-insulator-metal-insulator and insulator-metal-insulator (MIMI-IMI) interface with a 41\% overlap. (a) and (b) corresponds to the case of a metallic particle interacting with an optically thick and thin metallic film, respectively. Black solid curves show $|E_z|$ mode profiles of different SPPs. For these calculations, we used Palik \cite{palik1998handbook} silver for the metallic layers, refractive index of 1.4 for the insulator layers, dielectric gap sizes of 5 nm, and the thickness of the metallic substrate as semi-infinite for (a) and 10 nm for (b). (c) Dispersion relations of SPPs. (d) Mode overlap dispersion in the MIM-IM and MIMI-IMI interfaces.  }
  \label{SPP}
\end{figure}

Recently, gap plasmons \cite{esteban2010optical,moreau2012controlled,russell2012large,belacel2013controlling,rose2014control,akselrod2014probing,faggiani2015quenching} show their advantage in spontaneous emission enhancement for the corresponding more reliable control of the dielectric gap thinness. An optically thick metallic substrate is commonly used in previous reports \cite{russell2012large,belacel2013controlling,rose2014control,akselrod2014probing,faggiani2015quenching,pors2015quantum,lian2015efficient}, in order to obtain the highly-confined metal-insulator-metal (MIM) SPP within the dielectric gap. However, the thick film also induces large mode absorption, when the dielectric gap vanishes. Moreover, the QY of an emitter inside the gap is especially sensitive to its vertical position; the maximum QY is usually achieved if the emitter is placed at the center of the gap but becomes extremely low if the emitter is placed near metal.

To begin with, we show why optically thin metallic substrates can facilitate high-Purcell and high radiative-efficiency plasmonics via a mode-overlap analysis.
Film-coupled nanoparticles can be understood as Fabry-Perot cavities \cite{bozhevolnyi2007general,miyazaki2006squeezing,yang2012ultrasmall} of gap plasmons, with two radiative channels: one into propagating surface plasmon polaritons (SPPs), and another into photons via adiabatic tapering effect \cite{johnson2002adiabatic,yamamoto2010gap,fernandez2010collection} using nanoparticle edges.
Fig.~\ref{SPP}(a) shows the conventionally used metal-insulator-metal (MIM) SPP for emission enhancement.
If we reduce the thickness of metal substrate so that it is smaller than the skin depth of MIM SPP, the lower dielectric half space starts to have a decaying tail. We call this new type of SPP the metal-insulator-metal-insulator (MIMI) SPP [Fig.~\ref{SPP}(b)].
Surprisingly, although we use less metal, the MIMI SPP achieves better light confinement (smaller $\partial\omega/\partial k$) than the MIM SPP given the same frequency, as shown in the dispersion diagram [Fig.~\ref{SPP}(c)]. This indicates that the on-resonance local density of states of the MIMI SPP will be higher than that of the MIM SPP, if one replaces the top metal layer with a nanoparticle as a frequency-selecting cavity. A better mode overlap \cite{johnson2002adiabatic,palamaru2001photonic} (middle of Fig.~\ref{SPP}(a)(b) and see Supplementary Information) between the gap plasmon with the corresponding propagating SPP implies a larger radiative decay rate into propagating SPP than that in the case using an optically thick film. Fig.~\ref{SPP}(d) shows that the MIMI-IMI overlap is much larger than the MIM-IM overlap over a wide wavelength range, from near infrared to the entire visible spectrum.
%On the other hand, as the thickness of the metallic substrate becomes much smaller than the mean free path of electrons in silver ($\sim$50 nm at room temperature \cite{fuchs1938conductivity,davis2001interconnect}), mode absorption within the metallic film can be greatly suppressed.
Note that although the above analysis only discusses the mode matching between gap and propagating SPPs, the photon decay rate can be greatly enhanced via tapering the SPPs into photons using the momenta provided by nanoparticle edges, which we will show later.

\begin{figure}[hbtp]
  \centering
  \includegraphics[width=6 in]{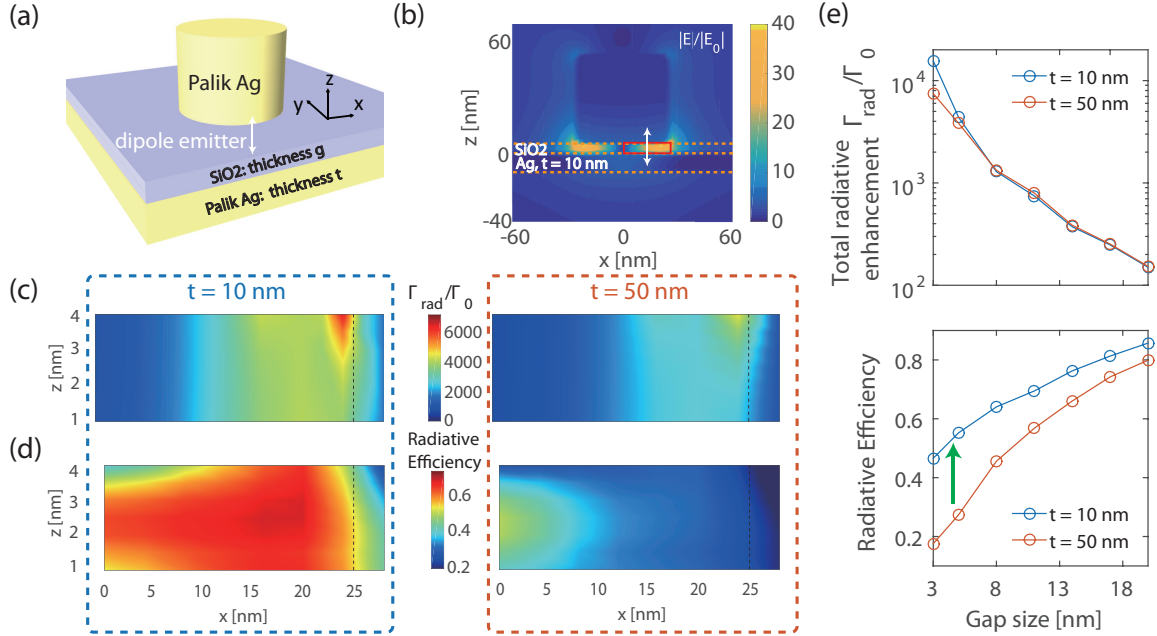}
  \caption{(a) Structure for spontaneous emission enhancement: a silver cylinder (diameter and height both 50 nm) sitting on top of a silver substrate (thickness $t$) and a dielectric (SiO$_2$, $n$=1.4) gap (thickness $g$). Free space refractive index is 1.4. (b) Normalized electric field $|E|\slash|E_0|$ of the gap plasmon resonance with $t = 10$ nm and $g=5$ nm. Electric field is mostly confined within the dielectric gap. The white arrow denotes a z-polarized dipole emitter and the red solid box defines the sweeping area of the dipole location. Orange dashed lines outline the interfaces between different layers. (c) Radiative enhancement and (d) radiative efficiency in the $x-z$ plane as a function of dipole location (left: $t = 10$ nm; right: $t = 50$ nm) with fixed $g=5$ nm. (e) Evolution of radiative enhancement (upper) and efficiency (lower) as a function of dielectric gap size $g$, with a thin ($t=10$ nm) and thick ($t=50$ nm) silver substrate. The green arrow indicates the increase of efficiency by decreasing substrate thickness. The size of the cylinder changes accordingly with different $g$ to maintain the resonance at $\sim$700 nm. The dipole stays at the center of the gap, and under the edge of the cylinder. }
  \label{emit1}
\end{figure}

Next, we move from the analytical modal analysis to rigorous computations of the enhanced emission characteristics for realistic structures.
We consider a structure with a silver cylinder on top of a silver thin film [Fig.~\ref{emit1}(a)]. The permittivities of the cylinder and the film are both Palik silver\cite{palik1998handbook} to offer a worst-case scenario analysis.
For this structure, the radiative (photon + plasmon) efficiency $\eta$ is calculated to be $\eta \sim$60\% and $\eta \sim$30\% for $t=10$ and $t=50$ nm respectively using the scattering and extinction cross-sections of the cylinder, as shown in Fig.~S5.
As the electric field is dominated by $E_z$, a z-polarized dipole (marked by the white arrow) is placed within the gap to probe the enhancement [Fig.~\ref{emit1}(b)]. A sweeping analysis of dipole location in the $x-z$ plane (marked by the solid red box) provides all the information about the enhancement due to the rotational symmetry of the structure.
As shown in Fig.~\ref{emit1}(c), the radiative decay rate $\Gamma_{\text {rad}}/\Gamma_0$ is generally higher with the thin film ($t=10$ nm) than that with the thick film ($t=50$ nm).
More surprisingly, $\eta$ in the $t=10$ nm case remains almost uniformly high in the $x-z$ plane with an average of $\sim$60\%, while that in the $t=50$ case drops to $\sim$30\% (Fig.~\ref{emit1}(d).
Both results are consistent with their scattering-extinction ratio (Fig.~S5).
Note that in the $t=10$ nm case, $\Gamma_{\text {rad}}/\Gamma_0$ remains high even for dipole locations within 1-nm distance from the metal surface, where absorption is always considered dominant \cite{akselrod2014probing,eggleston2015optical,anger2006enhancement}. If epitaxial silver is used for the metal substrate, similar results are obtained with even higher $\eta$, as shown in Fig.~S6.
Fig.~\ref{emit1}(e) compares $\Gamma_{\text {rad}}/\Gamma_0$ and $\eta$ as a function of dielectric gap size for $t=10$ and $t=50$ nm cases, with a fixed emitter at the center of the gap, and under the edge of the cylinder. The trends of $\Gamma_{\text {rad}}/\Gamma_0$ are similar. For $\eta$, in the $t=10$ nm case it remains higher for all gap sizes. The advantage becomes more striking with vanishing gap size (3$\sim$8 nm), where the thin substrate achieves a much higher enhancement and efficiency simultaneously.

The optically thin metallic susbstrates have two main advantages compared to the thick ones.
First, the cavity mode becomes less absorptive as shown by the loss per volume (smaller $\Gamma_{\text {abs}}$, see Fig.~S5). Second, the radiative decay rate is enhanced (larger $\Gamma_{\text {rad}}$) because of the improved mode overlap condition (Fig.~\ref{SPP}).

\begin{figure}[hbtp]
  \centering
  \includegraphics[width=3 in]{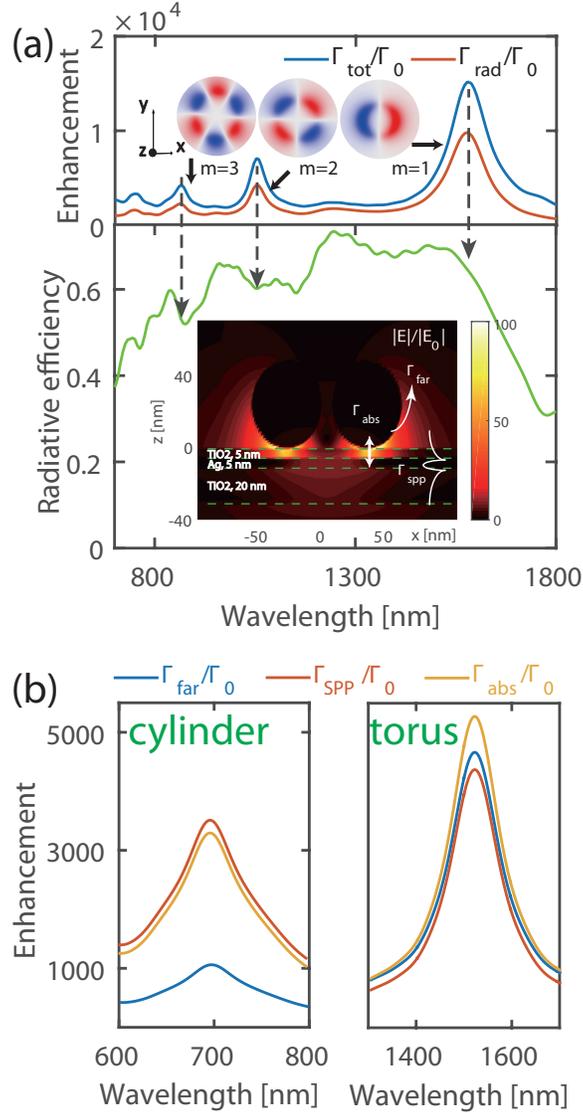}
  \caption{(a) Optically thin metal films enable high radiative efficiencies even for high-order (large-azimuthal-index, m) modes, which are typically less efficient in plasmonics. Emission enhancement and radiative efficiency of the torus-multifilm structure ($R=28$ nm, $r=24$ nm, and $t=5$ nm. Other configurations are the same as those defined in Fig.~\ref{sca1}.) are shown. Upper inset: E$_\text{z}$ profiles in the middle of the dielectric gap of the $m=1,2,3$ gap plasmon modes. Lower inset: Normalized electric field of the gap plasmon resonance of the torus with illustrated major decaying channels: free space radiation into the far field $\Gamma_{\text {far}}$, launched SPP $\Gamma_{\text {SPP}}$, and absorption $\Gamma_{\text {abs}}$ (including quenching and mode absorption). The white two-sided arrow indicates the location of the $z$-polarized dipole. Green dash lines denote the interfaces of different layers. (b) Radiation into surface plasmons can be converted to radiation in the far field by altering the nanoparticle shape, e.g., from a cylinder to a torus. The key to the large total ($\Gamma_{\text {far}}+\Gamma_{\text {SPP}}$) radiative emission, in either case, is the use of a thin-film metallic substrate.}
  \label{emit2}
\end{figure}

As there are two radiative channels in the gap plasmon structure (i.e., free space radiation into the far field $\Gamma_{\text {far}}$ and SPP excitation $\Gamma_{\text {SPP}}$), it is important to separate the total radiative decay rate $\Gamma_{\text {rad}}$ into the two channels (see supporting information) and know how to tailor their relative ratio.
It has been shown that tapered antennas (particles like spheres and tori) have higher radiative efficiencies than rigid antennas (particles like cubes and cylinders) \cite{faggiani2015quenching}. Here we show the ratio of $\Gamma_{\text {far}}$ and $\Gamma_{\text {SPP}}$ in the entire radiation can be tailored via the shape of nanoparticles.
We replace the cylinder with a torus, as shown in Fig.~\ref{emit2}. There are multiple orders of gap plasmon resonances (whispering gallery modes with the dielectric gap) in this structure. Usually the decay of high-order resonances of a plasmonic nanoantenna is dominated by absorption and thus are not very efficient for excitation or radiation.
However, with a thin metal substrate, the first three gap plasmon resonances of the structure (denoted by their azimuthal index $m$) all achieve considerably high enhancement, while maintaining high efficiencies [Fig.~\ref{emit2}(a)].
This result reveals the potential for high-efficiency harmonic generation and wave multiplexing.
For the cylinder, $\Gamma_{\text {SPP}}$ is the dominant radiative channel [Fig.~\ref{emit2}(b) left], making this structure an ideal candidate for a high excitation-efficiency plasmon source \cite{koenderink2009plasmon,gonzalez2011entanglement,gan2012proposal,kumar2014generation}.
While for the torus, $\Gamma_{\text {far}}$ is greatly boosted, which is useful for fluorescence applications \cite{akselrod2014probing,rose2014control,eggleston2015optical} [Fig.~\ref{emit2}(b) right].
Note that although the photon and plasmon excitation ratio is different in the two nanoparticles, it is the thin metallic substrate that gives rise to the high total radiative enhancement.

The aforementioned high-Q scattering and high-QY emission are deeply connected via the radiative efficiency $\eta$ but differ from each other. For scattering, $\text {FOM}_{\text {sca}}=Q/(1-\eta)$. For plasmon-enhanced emission, $\text {FOM}_{\text {emit}}\propto\eta/V$. Thus, two applications focus on $Q$ and $V$ respectively. What they need in common is a higher $\eta$ for either stronger scattering or higher quantum yield. Another difference is that a high-quality metallic substrate is not essential for high-efficiency (>50\%) emission (compare Fig.~\ref{emit1}(c)(d) with Fig.~S6), as the improved mode matching does not rely on low-absoprtion materials. Nevertheless, it is necessary if one intends to exceed the $Q_{\text {qs}}$ of the nanoparticle material by using the thin metallic substrate (see Fig.~\ref{sca2}).

It is also important to consider the practical feasibility of fabricating such high-quality thin films, and whether the material can be approximated with a local (bulk) permittivity.
%There may be concerns on the feasibility of fabricating high-quality metallic thin films from both theoretical and practical perspectives.
Theoretically, nonlocal effects \cite{mortensen2014generalized} induce additional loss when the dimension of plasmonic structures becomes small. Specifically for multifilms, the nonlocal effects are typically insignificant with geometrical sizes larger than 1$\sim$2 nm \cite{ciraci2012probing} (or $>\lambda_p/100$\cite{yan2012effects}, $\lambda_p$ is the plasma wavelength) in the gap plasmon resonances. In addition, the nanoparticles discussed in this Letter are generally large enough (size > 20 nm) such that the nonlocal effects are negligible, yet small enough (size < $\lambda$/10) such that the quasistatic approximation still holds. Overall, the local response approximation is still valid in the above analysis. Practically, the low-temperature epitaxial growth technique can provide a low growth rate (typically 1 angstrom/minute \cite{wu2014intrinsic}) while maintaining high film quality, making this technique ideal for the fabrication of low-loss ultrathin film ($\lesssim$ 10 nm).

In this letter, we show that optically thin metallic films offer an ideal platform for high-radiative-efficiency plasmonics. Using a thin metallic substrate, we achieve high-$Q$ and strong scattering that exceeds the quasistatic limit of the nanoparticle material.
Based on the improved mode matching condition, we predicted large-Purcell ($F_p>10^4$) and high-efficiency (>50\%) for gap-plasmon-enhanced spontaneous emission, maintained over the whole active region.
Future efforts can be made on particle designs that enable accurate and high dynamic-range control of the plasmon and photon excitation.
It will also be interesting to study how resonances interfere \cite{wang2006symmetry,mukherjee2010fanoshells,hsu2014theoretical} with each other on this platform.


\begin{acknowledgement}
The authors thank Prof. Koppens for the suggestion to look into epitaxially grown silver. The authors thank Di Zhu, Adi Pick, Dr. Homer Reid, Dr. Jianji Yang for helpful discussions.
Y. Y. was partly supported by the MRSEC Program of the
National Science Foundation under Grant No. DMR-1419807.
B. Z. and M. S. were partly supported by S3TEC, an Energy Frontier Research Center funded by the US Department of Energy under grant no. DE-SC0001299. B. Z. was partially supported by the United States-Israel Binational Science Foundation (BSF) under award no. 2013508. C. W. H. was partly supported by the National Science Foundation through grant no. DMR-1307632. O. D. M. was supported by the Army Research Office through the Institute for Soldier Nanotechnologies under contract no. W911NF-13-D-0001.
%
%Please use ``The authors thank \ldots'' rather than ``The
%authors would like to thank \ldots''.
%
%The author thanks Mats Dahlgren for version one of \textsf{achemso},
%and Donald Arseneau for the code taken from \textsf{cite} to move
%citations after punctuation. Many users have provided feedback on the
%class, which is reflected in all of the different demonstrations
%shown in this document.
%
\end{acknowledgement}

\begin{suppinfo}
Numerical methods (Supplementary Texts); Analytical mode overlap calculation (Supplementary Texts); Evolution of Mie and gap plasmon resonances when the torus is moving toward the multifilm (Fig.~S1); Multifilm transmission (Fig.~S2); Mode squeezing (Fig.~S3); Energy concentration coefficients as functions of the silver film thickness (Fig.~S4);  Radiation efficiencies of the silver cylinder and corresponding absorption per volume for thin and thick silver substrates (Fig.~S5); Spontaneous emission enhancement using epitaxial-grown silver substrates (Fig.S6).
\end{suppinfo}

\newpage

\bibliography{torusBib}

\providecommand{\latin}[1]{#1}
\providecommand*\mcitethebibliography{\thebibliography}
\csname @ifundefined\endcsname{endmcitethebibliography}
  {\let\endmcitethebibliography\endthebibliography}{}
\begin{mcitethebibliography}{70}
\providecommand*\natexlab[1]{#1}
\providecommand*\mciteSetBstSublistMode[1]{}
\providecommand*\mciteSetBstMaxWidthForm[2]{}
\providecommand*\mciteBstWouldAddEndPuncttrue
  {\def\EndOfBibitem{\unskip.}}
\providecommand*\mciteBstWouldAddEndPunctfalse
  {\let\EndOfBibitem\relax}
\providecommand*\mciteSetBstMidEndSepPunct[3]{}
\providecommand*\mciteSetBstSublistLabelBeginEnd[3]{}
\providecommand*\EndOfBibitem{}
\mciteSetBstSublistMode{f}
\mciteSetBstMaxWidthForm{subitem}{(\alph{mcitesubitemcount})}
\mciteSetBstSublistLabelBeginEnd
  {\mcitemaxwidthsubitemform\space}
  {\relax}
  {\relax}

\bibitem[Khurgin(2015)]{khurgin2015deal}
Khurgin,~J.~B. \emph{Nat. Nanotechnol.} \textbf{2015}, \emph{10}, 2--6\relax
\mciteBstWouldAddEndPuncttrue
\mciteSetBstMidEndSepPunct{\mcitedefaultmidpunct}
{\mcitedefaultendpunct}{\mcitedefaultseppunct}\relax
\EndOfBibitem
\bibitem[Sobhani \latin{et~al.}(2015)Sobhani, Manjavacas, Cao, McClain,
  Garc\'{i}a~de Abajo, Nordlander, and Halas]{sobhani2015pronounced}
Sobhani,~A.; Manjavacas,~A.; Cao,~Y.; McClain,~M.~J.; Garc\'{i}a~de
  Abajo,~F.~J.; Nordlander,~P.; Halas,~N.~J. \emph{Nano Lett.} \textbf{2015},
  \emph{15}, 6946--6951\relax
\mciteBstWouldAddEndPuncttrue
\mciteSetBstMidEndSepPunct{\mcitedefaultmidpunct}
{\mcitedefaultendpunct}{\mcitedefaultseppunct}\relax
\EndOfBibitem
\bibitem[Liu \latin{et~al.}(2012)Liu, Miroshnichenko, Neshev, and
  Kivshar]{liu2012broadband}
Liu,~W.; Miroshnichenko,~A.~E.; Neshev,~D.~N.; Kivshar,~Y.~S. \emph{ACS Nano}
  \textbf{2012}, \emph{6}, 5489--5497\relax
\mciteBstWouldAddEndPuncttrue
\mciteSetBstMidEndSepPunct{\mcitedefaultmidpunct}
{\mcitedefaultendpunct}{\mcitedefaultseppunct}\relax
\EndOfBibitem
\bibitem[Chang \latin{et~al.}(2011)Chang, Willingham, Slaughter, Khanal,
  Vigderman, Zubarev, and Link]{chang2011low}
Chang,~W.-S.; Willingham,~B.~A.; Slaughter,~L.~S.; Khanal,~B.~P.;
  Vigderman,~L.; Zubarev,~E.~R.; Link,~S. \emph{Proc. Natl. Acad. Sci.}
  \textbf{2011}, \emph{108}, 19879--19884\relax
\mciteBstWouldAddEndPuncttrue
\mciteSetBstMidEndSepPunct{\mcitedefaultmidpunct}
{\mcitedefaultendpunct}{\mcitedefaultseppunct}\relax
\EndOfBibitem
\bibitem[Peer \latin{et~al.}(2007)Peer, Karp, Hong, Farokhzad, Margalit, and
  Langer]{peer2007nanocarriers}
Peer,~D.; Karp,~J.~M.; Hong,~S.; Farokhzad,~O.~C.; Margalit,~R.; Langer,~R.
  \emph{Nat. Nanotechnol.} \textbf{2007}, \emph{2}, 751--760\relax
\mciteBstWouldAddEndPuncttrue
\mciteSetBstMidEndSepPunct{\mcitedefaultmidpunct}
{\mcitedefaultendpunct}{\mcitedefaultseppunct}\relax
\EndOfBibitem
\bibitem[Ota \latin{et~al.}(2013)Ota, Wang, Wang, Yin, and Zhang]{ota2013lipid}
Ota,~S.; Wang,~S.; Wang,~Y.; Yin,~X.; Zhang,~X. \emph{Nano Lett.}
  \textbf{2013}, \emph{13}, 2766--2770\relax
\mciteBstWouldAddEndPuncttrue
\mciteSetBstMidEndSepPunct{\mcitedefaultmidpunct}
{\mcitedefaultendpunct}{\mcitedefaultseppunct}\relax
\EndOfBibitem
\bibitem[Maier \latin{et~al.}(2003)Maier, Kik, Atwater, Meltzer, Harel, Koel,
  and Requicha]{maier2003local}
Maier,~S.~A.; Kik,~P.~G.; Atwater,~H.~A.; Meltzer,~S.; Harel,~E.; Koel,~B.~E.;
  Requicha,~A.~A. \emph{Nat. Mater.} \textbf{2003}, \emph{2}, 229--232\relax
\mciteBstWouldAddEndPuncttrue
\mciteSetBstMidEndSepPunct{\mcitedefaultmidpunct}
{\mcitedefaultendpunct}{\mcitedefaultseppunct}\relax
\EndOfBibitem
\bibitem[Wu \latin{et~al.}(2014)Wu, Zhang, Estakhri, Zhao, Kim, Zhang, Liu,
  Pribil, Al{\`u}, Shih, and Li]{wu2014intrinsic}
Wu,~Y.; Zhang,~C.; Estakhri,~N.~M.; Zhao,~Y.; Kim,~J.; Zhang,~M.; Liu,~X.-X.;
  Pribil,~G.~K.; Al{\`u},~A.; Shih,~C.-K.; Li,~X. \emph{Adv. Mater.}
  \textbf{2014}, \emph{26}, 6106--6110\relax
\mciteBstWouldAddEndPuncttrue
\mciteSetBstMidEndSepPunct{\mcitedefaultmidpunct}
{\mcitedefaultendpunct}{\mcitedefaultseppunct}\relax
\EndOfBibitem
\bibitem[McPeak \latin{et~al.}(2015)McPeak, Jayanti, Kress, Meyer, Iotti,
  Rossinelli, and Norris]{mcpeak2015plasmonic}
McPeak,~K.~M.; Jayanti,~S.~V.; Kress,~S.~J.; Meyer,~S.; Iotti,~S.;
  Rossinelli,~A.; Norris,~D.~J. \emph{ACS Photonics} \textbf{2015}, \emph{2},
  326--333\relax
\mciteBstWouldAddEndPuncttrue
\mciteSetBstMidEndSepPunct{\mcitedefaultmidpunct}
{\mcitedefaultendpunct}{\mcitedefaultseppunct}\relax
\EndOfBibitem
\bibitem[Babar and Weaver(2015)Babar, and Weaver]{babar2015optical}
Babar,~S.; Weaver,~J. \emph{Appl. Opt.} \textbf{2015}, \emph{54},
  477--481\relax
\mciteBstWouldAddEndPuncttrue
\mciteSetBstMidEndSepPunct{\mcitedefaultmidpunct}
{\mcitedefaultendpunct}{\mcitedefaultseppunct}\relax
\EndOfBibitem
\bibitem[Miller \latin{et~al.}(2014)Miller, Johnson, and
  Rodriguez]{miller2014effectiveness}
Miller,~O.~D.; Johnson,~S.~G.; Rodriguez,~A.~W. \emph{Phys. Rev. Lett.}
  \textbf{2014}, \emph{112}, 157402\relax
\mciteBstWouldAddEndPuncttrue
\mciteSetBstMidEndSepPunct{\mcitedefaultmidpunct}
{\mcitedefaultendpunct}{\mcitedefaultseppunct}\relax
\EndOfBibitem
\bibitem[Koenderink(2010)]{koenderink2010use}
Koenderink,~A.~F. \emph{Opt. Lett.} \textbf{2010}, \emph{35}, 4208--4210\relax
\mciteBstWouldAddEndPuncttrue
\mciteSetBstMidEndSepPunct{\mcitedefaultmidpunct}
{\mcitedefaultendpunct}{\mcitedefaultseppunct}\relax
\EndOfBibitem
\bibitem[Sauvan \latin{et~al.}(2013)Sauvan, Hugonin, Maksymov, and
  Lalanne]{sauvan2013theory}
Sauvan,~C.; Hugonin,~J.-P.; Maksymov,~I.; Lalanne,~P. \emph{Phys. Rev. Lett.}
  \textbf{2013}, \emph{110}, 237401\relax
\mciteBstWouldAddEndPuncttrue
\mciteSetBstMidEndSepPunct{\mcitedefaultmidpunct}
{\mcitedefaultendpunct}{\mcitedefaultseppunct}\relax
\EndOfBibitem
\bibitem[Kristensen and Hughes(2013)Kristensen, and
  Hughes]{kristensen2013modes}
Kristensen,~P.~T.; Hughes,~S. \emph{ACS Photonics} \textbf{2013}, \emph{1},
  2--10\relax
\mciteBstWouldAddEndPuncttrue
\mciteSetBstMidEndSepPunct{\mcitedefaultmidpunct}
{\mcitedefaultendpunct}{\mcitedefaultseppunct}\relax
\EndOfBibitem
\bibitem[Lee and El-Sayed(2005)Lee, and El-Sayed]{lee2005dependence}
Lee,~K.-S.; El-Sayed,~M.~A. \emph{J. Phys. Chem. B} \textbf{2005}, \emph{109},
  20331--20338\relax
\mciteBstWouldAddEndPuncttrue
\mciteSetBstMidEndSepPunct{\mcitedefaultmidpunct}
{\mcitedefaultendpunct}{\mcitedefaultseppunct}\relax
\EndOfBibitem
\bibitem[El-Sayed \latin{et~al.}(2005)El-Sayed, Huang, and
  El-Sayed]{el2005surface}
El-Sayed,~I.~H.; Huang,~X.; El-Sayed,~M.~A. \emph{Nano Lett.} \textbf{2005},
  \emph{5}, 829--834\relax
\mciteBstWouldAddEndPuncttrue
\mciteSetBstMidEndSepPunct{\mcitedefaultmidpunct}
{\mcitedefaultendpunct}{\mcitedefaultseppunct}\relax
\EndOfBibitem
\bibitem[Anker \latin{et~al.}(2008)Anker, Hall, Lyandres, Shah, Zhao, and
  Van~Duyne]{anker2008biosensing}
Anker,~J.~N.; Hall,~W.~P.; Lyandres,~O.; Shah,~N.~C.; Zhao,~J.;
  Van~Duyne,~R.~P. \emph{Nat. Mater.} \textbf{2008}, \emph{7}, 442--453\relax
\mciteBstWouldAddEndPuncttrue
\mciteSetBstMidEndSepPunct{\mcitedefaultmidpunct}
{\mcitedefaultendpunct}{\mcitedefaultseppunct}\relax
\EndOfBibitem
\bibitem[Saha \latin{et~al.}(2012)Saha, Agasti, Kim, Li, and
  Rotello]{saha2012gold}
Saha,~K.; Agasti,~S.~S.; Kim,~C.; Li,~X.; Rotello,~V.~M. \emph{Chem. Rev.}
  \textbf{2012}, \emph{112}, 2739--2779\relax
\mciteBstWouldAddEndPuncttrue
\mciteSetBstMidEndSepPunct{\mcitedefaultmidpunct}
{\mcitedefaultendpunct}{\mcitedefaultseppunct}\relax
\EndOfBibitem
\bibitem[Hsu \latin{et~al.}(2014)Hsu, Zhen, Qiu, Shapira, DeLacy, Joannopoulos,
  and Solja{\v{c}}i{\'c}]{hsu2014transparent}
Hsu,~C.~W.; Zhen,~B.; Qiu,~W.; Shapira,~O.; DeLacy,~B.~G.; Joannopoulos,~J.~D.;
  Solja{\v{c}}i{\'c},~M. \emph{Nat. Commun.} \textbf{2014}, \emph{5},
  3152\relax
\mciteBstWouldAddEndPuncttrue
\mciteSetBstMidEndSepPunct{\mcitedefaultmidpunct}
{\mcitedefaultendpunct}{\mcitedefaultseppunct}\relax
\EndOfBibitem
\bibitem[Hsu \latin{et~al.}(2015)Hsu, Miller, Johnson, and
  Solja{\v{c}}i{\'c}]{hsu2015optimization}
Hsu,~C.~W.; Miller,~O.~D.; Johnson,~S.~G.; Solja{\v{c}}i{\'c},~M. \emph{Opt.
  Express} \textbf{2015}, \emph{23}, 9516--9526\relax
\mciteBstWouldAddEndPuncttrue
\mciteSetBstMidEndSepPunct{\mcitedefaultmidpunct}
{\mcitedefaultendpunct}{\mcitedefaultseppunct}\relax
\EndOfBibitem
\bibitem[Saito and Tatsuma(2015)Saito, and Tatsuma]{saito2015transparent}
Saito,~K.; Tatsuma,~T. \emph{Nanoscale} \textbf{2015}, \emph{7},
  20365--20368\relax
\mciteBstWouldAddEndPuncttrue
\mciteSetBstMidEndSepPunct{\mcitedefaultmidpunct}
{\mcitedefaultendpunct}{\mcitedefaultseppunct}\relax
\EndOfBibitem
\bibitem[Wang and Shen(2006)Wang, and Shen]{PhysRevLett.97.206806}
Wang,~F.; Shen,~Y.~R. \emph{Phys. Rev. Lett.} \textbf{2006}, \emph{97},
  206806\relax
\mciteBstWouldAddEndPuncttrue
\mciteSetBstMidEndSepPunct{\mcitedefaultmidpunct}
{\mcitedefaultendpunct}{\mcitedefaultseppunct}\relax
\EndOfBibitem
\bibitem[Raman \latin{et~al.}(2013)Raman, Shin, and
  Fan]{PhysRevLett.110.183901}
Raman,~A.; Shin,~W.; Fan,~S. \emph{Phys. Rev. Lett.} \textbf{2013}, \emph{110},
  183901\relax
\mciteBstWouldAddEndPuncttrue
\mciteSetBstMidEndSepPunct{\mcitedefaultmidpunct}
{\mcitedefaultendpunct}{\mcitedefaultseppunct}\relax
\EndOfBibitem
\bibitem[Miller \latin{et~al.}(2014)Miller, Hsu, Reid, Qiu, DeLacy,
  Joannopoulos, Solja\ifmmode \check{c}\else \v{c}\fi{}i\ifmmode~\acute{c}\else
  \'{c}\fi{}, and Johnson]{PhysRevLett.112.123903}
Miller,~O.~D.; Hsu,~C.~W.; Reid,~M. T.~H.; Qiu,~W.; DeLacy,~B.~G.;
  Joannopoulos,~J.~D.; Solja\ifmmode \check{c}\else
  \v{c}\fi{}i\ifmmode~\acute{c}\else \'{c}\fi{},~M.; Johnson,~S.~G. \emph{Phys.
  Rev. Lett.} \textbf{2014}, \emph{112}, 123903\relax
\mciteBstWouldAddEndPuncttrue
\mciteSetBstMidEndSepPunct{\mcitedefaultmidpunct}
{\mcitedefaultendpunct}{\mcitedefaultseppunct}\relax
\EndOfBibitem
\bibitem[Miller \latin{et~al.}(2016)Miller, Polimeridis, Reid, Hsu, DeLacy,
  Joannopoulos, Solja\v{c}i\'{c}, and Johnson]{sigmaLimitArxiv}
Miller,~O.~D.; Polimeridis,~A.~G.; Reid,~M. T.~H.; Hsu,~C.~W.; DeLacy,~B.~G.;
  Joannopoulos,~J.~D.; Solja\v{c}i\'{c},~M.; Johnson,~S.~G. \emph{Opt. Express}
  \textbf{2016}, \emph{24}, 3329--3364\relax
\mciteBstWouldAddEndPuncttrue
\mciteSetBstMidEndSepPunct{\mcitedefaultmidpunct}
{\mcitedefaultendpunct}{\mcitedefaultseppunct}\relax
\EndOfBibitem
\bibitem[Anger \latin{et~al.}(2006)Anger, Bharadwaj, and
  Novotny]{anger2006enhancement}
Anger,~P.; Bharadwaj,~P.; Novotny,~L. \emph{Phys. Rev. Lett.} \textbf{2006},
  \emph{96}, 113002\relax
\mciteBstWouldAddEndPuncttrue
\mciteSetBstMidEndSepPunct{\mcitedefaultmidpunct}
{\mcitedefaultendpunct}{\mcitedefaultseppunct}\relax
\EndOfBibitem
\bibitem[K{\"u}hn \latin{et~al.}(2006)K{\"u}hn, H{\aa}kanson, Rogobete, and
  Sandoghdar]{kuhn2006enhancement}
K{\"u}hn,~S.; H{\aa}kanson,~U.; Rogobete,~L.; Sandoghdar,~V. \emph{Phys. Rev.
  Lett.} \textbf{2006}, \emph{97}, 017402\relax
\mciteBstWouldAddEndPuncttrue
\mciteSetBstMidEndSepPunct{\mcitedefaultmidpunct}
{\mcitedefaultendpunct}{\mcitedefaultseppunct}\relax
\EndOfBibitem
\bibitem[Russell \latin{et~al.}(2012)Russell, Liu, Cui, and
  Hu]{russell2012large}
Russell,~K.~J.; Liu,~T.-L.; Cui,~S.; Hu,~E.~L. \emph{Nat. Photonics}
  \textbf{2012}, \emph{6}, 459--462\relax
\mciteBstWouldAddEndPuncttrue
\mciteSetBstMidEndSepPunct{\mcitedefaultmidpunct}
{\mcitedefaultendpunct}{\mcitedefaultseppunct}\relax
\EndOfBibitem
\bibitem[Rose \latin{et~al.}(2014)Rose, Hoang, McGuire, Mock, Cirac{\`\i},
  Smith, and Mikkelsen]{rose2014control}
Rose,~A.; Hoang,~T.~B.; McGuire,~F.; Mock,~J.~J.; Cirac{\`\i},~C.;
  Smith,~D.~R.; Mikkelsen,~M.~H. \emph{Nano Lett.} \textbf{2014}, \emph{14},
  4797--4802\relax
\mciteBstWouldAddEndPuncttrue
\mciteSetBstMidEndSepPunct{\mcitedefaultmidpunct}
{\mcitedefaultendpunct}{\mcitedefaultseppunct}\relax
\EndOfBibitem
\bibitem[Akselrod \latin{et~al.}(2014)Akselrod, Argyropoulos, Hoang,
  Cirac{\`\i}, Fang, Huang, Smith, and Mikkelsen]{akselrod2014probing}
Akselrod,~G.~M.; Argyropoulos,~C.; Hoang,~T.~B.; Cirac{\`\i},~C.; Fang,~C.;
  Huang,~J.; Smith,~D.~R.; Mikkelsen,~M.~H. \emph{Nat. Photonics}
  \textbf{2014}, \emph{8}, 835--840\relax
\mciteBstWouldAddEndPuncttrue
\mciteSetBstMidEndSepPunct{\mcitedefaultmidpunct}
{\mcitedefaultendpunct}{\mcitedefaultseppunct}\relax
\EndOfBibitem
\bibitem[Eggleston \latin{et~al.}(2015)Eggleston, Messer, Zhang, Yablonovitch,
  and Wu]{eggleston2015optical}
Eggleston,~M.~S.; Messer,~K.; Zhang,~L.; Yablonovitch,~E.; Wu,~M.~C.
  \emph{Proc. Natl. Acad. Sci.} \textbf{2015}, \emph{112}, 1704--1709\relax
\mciteBstWouldAddEndPuncttrue
\mciteSetBstMidEndSepPunct{\mcitedefaultmidpunct}
{\mcitedefaultendpunct}{\mcitedefaultseppunct}\relax
\EndOfBibitem
\bibitem[Pelton(2015)]{pelton2015modified}
Pelton,~M. \emph{Nat. Photonics} \textbf{2015}, \emph{9}, 427--435\relax
\mciteBstWouldAddEndPuncttrue
\mciteSetBstMidEndSepPunct{\mcitedefaultmidpunct}
{\mcitedefaultendpunct}{\mcitedefaultseppunct}\relax
\EndOfBibitem
\bibitem[Purcell(1946)]{purcell1946spontaneous}
Purcell,~E.~M. \emph{Phys. Rev.} \textbf{1946}, \emph{69}, 681\relax
\mciteBstWouldAddEndPuncttrue
\mciteSetBstMidEndSepPunct{\mcitedefaultmidpunct}
{\mcitedefaultendpunct}{\mcitedefaultseppunct}\relax
\EndOfBibitem
\bibitem[Vesseur \latin{et~al.}(2010)Vesseur, de~Abajo, and
  Polman]{vesseur2010broadband}
Vesseur,~E. J.~R.; de~Abajo,~F. J.~G.; Polman,~A. \emph{Phys. Rev. B}
  \textbf{2010}, \emph{82}, 165419\relax
\mciteBstWouldAddEndPuncttrue
\mciteSetBstMidEndSepPunct{\mcitedefaultmidpunct}
{\mcitedefaultendpunct}{\mcitedefaultseppunct}\relax
\EndOfBibitem
\bibitem[Bohren and Huffman(2008)Bohren, and Huffman]{bohren2008absorption}
Bohren,~C.~F.; Huffman,~D.~R. \emph{Absorption and scattering of light by small
  particles}; John Wiley \& Sons, 2008\relax
\mciteBstWouldAddEndPuncttrue
\mciteSetBstMidEndSepPunct{\mcitedefaultmidpunct}
{\mcitedefaultendpunct}{\mcitedefaultseppunct}\relax
\EndOfBibitem
\bibitem[Faggiani \latin{et~al.}(2015)Faggiani, Yang, and
  Lalanne]{faggiani2015quenching}
Faggiani,~R.; Yang,~J.; Lalanne,~P. \emph{ACS Photonics} \textbf{2015},
  \emph{2}, 1739--1744\relax
\mciteBstWouldAddEndPuncttrue
\mciteSetBstMidEndSepPunct{\mcitedefaultmidpunct}
{\mcitedefaultendpunct}{\mcitedefaultseppunct}\relax
\EndOfBibitem
\bibitem[Novotny and Hecht(2012)Novotny, and Hecht]{novotny2012principles}
Novotny,~L.; Hecht,~B. \emph{Principles of nano-optics}; Cambridge university
  press, 2012\relax
\mciteBstWouldAddEndPuncttrue
\mciteSetBstMidEndSepPunct{\mcitedefaultmidpunct}
{\mcitedefaultendpunct}{\mcitedefaultseppunct}\relax
\EndOfBibitem
\bibitem[Esteban \latin{et~al.}(2010)Esteban, Teperik, and
  Greffet]{esteban2010optical}
Esteban,~R.; Teperik,~T.; Greffet,~J.-J. \emph{Phys. Rev. Lett.} \textbf{2010},
  \emph{104}, 026802\relax
\mciteBstWouldAddEndPuncttrue
\mciteSetBstMidEndSepPunct{\mcitedefaultmidpunct}
{\mcitedefaultendpunct}{\mcitedefaultseppunct}\relax
\EndOfBibitem
\bibitem[Moreau \latin{et~al.}(2012)Moreau, Cirac{\`\i}, Mock, Hill, Wang,
  Wiley, Chilkoti, and Smith]{moreau2012controlled}
Moreau,~A.; Cirac{\`\i},~C.; Mock,~J.~J.; Hill,~R.~T.; Wang,~Q.; Wiley,~B.~J.;
  Chilkoti,~A.; Smith,~D.~R. \emph{Nature} \textbf{2012}, \emph{492},
  86--89\relax
\mciteBstWouldAddEndPuncttrue
\mciteSetBstMidEndSepPunct{\mcitedefaultmidpunct}
{\mcitedefaultendpunct}{\mcitedefaultseppunct}\relax
\EndOfBibitem
\bibitem[Belacel \latin{et~al.}(2013)Belacel, Habert, Bigourdan, Marquier,
  Hugonin, de~Vasconcellos, Lafosse, Coolen, Schwob, Javaux, Dubertret,
  Greffet, Senellart, and Maitre]{belacel2013controlling}
Belacel,~C.; Habert,~B.; Bigourdan,~F.; Marquier,~F.; Hugonin,~J.-P.;
  de~Vasconcellos,~S.~M.; Lafosse,~X.; Coolen,~L.; Schwob,~C.; Javaux,~C.;
  Dubertret,~B.; Greffet,~J.-J.; Senellart,~P.; Maitre,~A. \emph{Nano Lett.}
  \textbf{2013}, \emph{13}, 1516--1521\relax
\mciteBstWouldAddEndPuncttrue
\mciteSetBstMidEndSepPunct{\mcitedefaultmidpunct}
{\mcitedefaultendpunct}{\mcitedefaultseppunct}\relax
\EndOfBibitem
\bibitem[Koenderink(2009)]{koenderink2009plasmon}
Koenderink,~A.~F. \emph{Nano Lett.} \textbf{2009}, \emph{9}, 4228--4233\relax
\mciteBstWouldAddEndPuncttrue
\mciteSetBstMidEndSepPunct{\mcitedefaultmidpunct}
{\mcitedefaultendpunct}{\mcitedefaultseppunct}\relax
\EndOfBibitem
\bibitem[Gonzalez-Tudela \latin{et~al.}(2011)Gonzalez-Tudela, Martin-Cano,
  Moreno, Martin-Moreno, Tejedor, and Garcia-Vidal]{gonzalez2011entanglement}
Gonzalez-Tudela,~A.; Martin-Cano,~D.; Moreno,~E.; Martin-Moreno,~L.;
  Tejedor,~C.; Garcia-Vidal,~F.~J. \emph{Phys. Rev. Lett.} \textbf{2011},
  \emph{106}, 020501\relax
\mciteBstWouldAddEndPuncttrue
\mciteSetBstMidEndSepPunct{\mcitedefaultmidpunct}
{\mcitedefaultendpunct}{\mcitedefaultseppunct}\relax
\EndOfBibitem
\bibitem[Gan \latin{et~al.}(2012)Gan, Hugonin, and Lalanne]{gan2012proposal}
Gan,~C.~H.; Hugonin,~J.-P.; Lalanne,~P. \emph{Phys. Rev. X} \textbf{2012},
  \emph{2}, 021008\relax
\mciteBstWouldAddEndPuncttrue
\mciteSetBstMidEndSepPunct{\mcitedefaultmidpunct}
{\mcitedefaultendpunct}{\mcitedefaultseppunct}\relax
\EndOfBibitem
\bibitem[Kumar \latin{et~al.}(2014)Kumar, Kristiansen, Huck, and
  Andersen]{kumar2014generation}
Kumar,~S.; Kristiansen,~N.~I.; Huck,~A.; Andersen,~U.~L. \emph{Nano Lett.}
  \textbf{2014}, \emph{14}, 663--669\relax
\mciteBstWouldAddEndPuncttrue
\mciteSetBstMidEndSepPunct{\mcitedefaultmidpunct}
{\mcitedefaultendpunct}{\mcitedefaultseppunct}\relax
\EndOfBibitem
\bibitem[Newton(2013)]{newton2013scattering}
Newton,~R.~G. \emph{Scattering theory of waves and particles}; Springer Science
  \& Business Media, 2013\relax
\mciteBstWouldAddEndPuncttrue
\mciteSetBstMidEndSepPunct{\mcitedefaultmidpunct}
{\mcitedefaultendpunct}{\mcitedefaultseppunct}\relax
\EndOfBibitem
\bibitem[Hamam \latin{et~al.}(2007)Hamam, Karalis, Joannopoulos, and
  Solja\ifmmode \check{c}\else \v{c}\fi{}i\ifmmode~\acute{c}\else
  \'{c}\fi{}]{PhysRevA.75.053801}
Hamam,~R.~E.; Karalis,~A.; Joannopoulos,~J.~D.; Solja\ifmmode \check{c}\else
  \v{c}\fi{}i\ifmmode~\acute{c}\else \'{c}\fi{},~M. \emph{Phys. Rev. A}
  \textbf{2007}, \emph{75}, 053801\relax
\mciteBstWouldAddEndPuncttrue
\mciteSetBstMidEndSepPunct{\mcitedefaultmidpunct}
{\mcitedefaultendpunct}{\mcitedefaultseppunct}\relax
\EndOfBibitem
\bibitem[Ruan and Fan(2012)Ruan, and Fan]{PhysRevA.85.043828}
Ruan,~Z.; Fan,~S. \emph{Phys. Rev. A} \textbf{2012}, \emph{85}, 043828\relax
\mciteBstWouldAddEndPuncttrue
\mciteSetBstMidEndSepPunct{\mcitedefaultmidpunct}
{\mcitedefaultendpunct}{\mcitedefaultseppunct}\relax
\EndOfBibitem
\bibitem[Mary \latin{et~al.}(2005)Mary, Dereux, and Ferrell]{mary2005localized}
Mary,~A.; Dereux,~A.; Ferrell,~T.~L. \emph{Phys. Rev. B} \textbf{2005},
  \emph{72}, 155426\relax
\mciteBstWouldAddEndPuncttrue
\mciteSetBstMidEndSepPunct{\mcitedefaultmidpunct}
{\mcitedefaultendpunct}{\mcitedefaultseppunct}\relax
\EndOfBibitem
\bibitem[Dutta \latin{et~al.}(2008)Dutta, Ali, Brandl, Park, and
  Nordlander]{dutta2008plasmonic}
Dutta,~C.~M.; Ali,~T.~A.; Brandl,~D.~W.; Park,~T.-H.; Nordlander,~P. \emph{J.
  Chem. Phys.} \textbf{2008}, \emph{129}, 084706\relax
\mciteBstWouldAddEndPuncttrue
\mciteSetBstMidEndSepPunct{\mcitedefaultmidpunct}
{\mcitedefaultendpunct}{\mcitedefaultseppunct}\relax
\EndOfBibitem
\bibitem[Teperik and Degiron(2011)Teperik, and Degiron]{teperik2011numerical}
Teperik,~T.; Degiron,~A. \emph{Phys. Rev. B} \textbf{2011}, \emph{83},
  245408\relax
\mciteBstWouldAddEndPuncttrue
\mciteSetBstMidEndSepPunct{\mcitedefaultmidpunct}
{\mcitedefaultendpunct}{\mcitedefaultseppunct}\relax
\EndOfBibitem
\bibitem[Rakovich \latin{et~al.}(2015)Rakovich, Albella, and
  Maier]{rakovich2015plasmonic}
Rakovich,~A.; Albella,~P.; Maier,~S.~A. \emph{ACS Nano} \textbf{2015},
  \emph{9}, 2648--2658\relax
\mciteBstWouldAddEndPuncttrue
\mciteSetBstMidEndSepPunct{\mcitedefaultmidpunct}
{\mcitedefaultendpunct}{\mcitedefaultseppunct}\relax
\EndOfBibitem
\bibitem[Palik(1998)]{palik1998handbook}
Palik,~E.~D. \emph{Handbook of optical constants of solids}; Academic press,
  1998; Vol.~3\relax
\mciteBstWouldAddEndPuncttrue
\mciteSetBstMidEndSepPunct{\mcitedefaultmidpunct}
{\mcitedefaultendpunct}{\mcitedefaultseppunct}\relax
\EndOfBibitem
\bibitem[Kim(1996)]{kim1996simultaneous}
Kim,~S. \emph{Appl. Opt.} \textbf{1996}, \emph{35}, 6703--6707\relax
\mciteBstWouldAddEndPuncttrue
\mciteSetBstMidEndSepPunct{\mcitedefaultmidpunct}
{\mcitedefaultendpunct}{\mcitedefaultseppunct}\relax
\EndOfBibitem
\bibitem[Yamamoto \latin{et~al.}(2010)Yamamoto, Ohtani, and Garcia~de
  Abajo]{yamamoto2010gap}
Yamamoto,~N.; Ohtani,~S.; Garcia~de Abajo,~F.~J. \emph{Nano Lett.}
  \textbf{2010}, \emph{11}, 91--95\relax
\mciteBstWouldAddEndPuncttrue
\mciteSetBstMidEndSepPunct{\mcitedefaultmidpunct}
{\mcitedefaultendpunct}{\mcitedefaultseppunct}\relax
\EndOfBibitem
\bibitem[Ruppin(2002)]{ruppin2002electromagnetic}
Ruppin,~R. \emph{Phys. Lett. A} \textbf{2002}, \emph{299}, 309--312\relax
\mciteBstWouldAddEndPuncttrue
\mciteSetBstMidEndSepPunct{\mcitedefaultmidpunct}
{\mcitedefaultendpunct}{\mcitedefaultseppunct}\relax
\EndOfBibitem
\bibitem[Hsu \latin{et~al.}(2014)Hsu, DeLacy, Johnson, Joannopoulos, and
  Solja\ifmmode \check{c}\else \v{c}\fi{}i\ifmmode~\acute{c}\else
  \'{c}\fi{}]{hsu2014theoretical}
Hsu,~C.~W.; DeLacy,~B.~G.; Johnson,~S.~G.; Joannopoulos,~J.~D.; Solja\ifmmode
  \check{c}\else \v{c}\fi{}i\ifmmode~\acute{c}\else \'{c}\fi{},~M. \emph{Nano
  letters} \textbf{2014}, \emph{14}, 2783--2788\relax
\mciteBstWouldAddEndPuncttrue
\mciteSetBstMidEndSepPunct{\mcitedefaultmidpunct}
{\mcitedefaultendpunct}{\mcitedefaultseppunct}\relax
\EndOfBibitem
\bibitem[Pors and Bozhevolnyi(2015)Pors, and Bozhevolnyi]{pors2015quantum}
Pors,~A.; Bozhevolnyi,~S.~I. \emph{ACS Photonics} \textbf{2015}, \emph{2},
  228--236\relax
\mciteBstWouldAddEndPuncttrue
\mciteSetBstMidEndSepPunct{\mcitedefaultmidpunct}
{\mcitedefaultendpunct}{\mcitedefaultseppunct}\relax
\EndOfBibitem
\bibitem[Lian \latin{et~al.}(2015)Lian, Gu, Ren, Zhang, Wang, and
  Gong]{lian2015efficient}
Lian,~H.; Gu,~Y.; Ren,~J.; Zhang,~F.; Wang,~L.; Gong,~Q. \emph{Phys. Rev.
  Lett.} \textbf{2015}, \emph{114}, 193002\relax
\mciteBstWouldAddEndPuncttrue
\mciteSetBstMidEndSepPunct{\mcitedefaultmidpunct}
{\mcitedefaultendpunct}{\mcitedefaultseppunct}\relax
\EndOfBibitem
\bibitem[Bozhevolnyi and S{\o}ndergaard(2007)Bozhevolnyi, and
  S{\o}ndergaard]{bozhevolnyi2007general}
Bozhevolnyi,~S.~I.; S{\o}ndergaard,~T. \emph{Opt. Express} \textbf{2007},
  \emph{15}, 10869--10877\relax
\mciteBstWouldAddEndPuncttrue
\mciteSetBstMidEndSepPunct{\mcitedefaultmidpunct}
{\mcitedefaultendpunct}{\mcitedefaultseppunct}\relax
\EndOfBibitem
\bibitem[Miyazaki and Kurokawa(2006)Miyazaki, and
  Kurokawa]{miyazaki2006squeezing}
Miyazaki,~H.~T.; Kurokawa,~Y. \emph{Phys. Rev. Lett.} \textbf{2006}, \emph{96},
  097401\relax
\mciteBstWouldAddEndPuncttrue
\mciteSetBstMidEndSepPunct{\mcitedefaultmidpunct}
{\mcitedefaultendpunct}{\mcitedefaultseppunct}\relax
\EndOfBibitem
\bibitem[Yang \latin{et~al.}(2012)Yang, Sauvan, Jouanin, Collin, Pelouard, and
  Lalanne]{yang2012ultrasmall}
Yang,~J.; Sauvan,~C.; Jouanin,~A.; Collin,~S.; Pelouard,~J.-L.; Lalanne,~P.
  \emph{Opt. Express} \textbf{2012}, \emph{20}, 16880--16891\relax
\mciteBstWouldAddEndPuncttrue
\mciteSetBstMidEndSepPunct{\mcitedefaultmidpunct}
{\mcitedefaultendpunct}{\mcitedefaultseppunct}\relax
\EndOfBibitem
\bibitem[Johnson \latin{et~al.}(2002)Johnson, Bienstman, Skorobogatiy,
  Ibanescu, Lidorikis, and Joannopoulos]{johnson2002adiabatic}
Johnson,~S.~G.; Bienstman,~P.; Skorobogatiy,~M.; Ibanescu,~M.; Lidorikis,~E.;
  Joannopoulos,~J. \emph{Phys. Rev. E} \textbf{2002}, \emph{66}, 066608\relax
\mciteBstWouldAddEndPuncttrue
\mciteSetBstMidEndSepPunct{\mcitedefaultmidpunct}
{\mcitedefaultendpunct}{\mcitedefaultseppunct}\relax
\EndOfBibitem
\bibitem[Fern{\'a}ndez-Dom{\'\i}nguez
  \latin{et~al.}(2010)Fern{\'a}ndez-Dom{\'\i}nguez, Maier, and
  Pendry]{fernandez2010collection}
Fern{\'a}ndez-Dom{\'\i}nguez,~A.; Maier,~S.; Pendry,~J. \emph{Phys. Rev. Lett.}
  \textbf{2010}, \emph{105}, 266807\relax
\mciteBstWouldAddEndPuncttrue
\mciteSetBstMidEndSepPunct{\mcitedefaultmidpunct}
{\mcitedefaultendpunct}{\mcitedefaultseppunct}\relax
\EndOfBibitem
\bibitem[Palamaru and Lalanne(2001)Palamaru, and Lalanne]{palamaru2001photonic}
Palamaru,~M.; Lalanne,~P. \emph{Appl. Phys. Lett.} \textbf{2001}, \emph{78},
  1466--1468\relax
\mciteBstWouldAddEndPuncttrue
\mciteSetBstMidEndSepPunct{\mcitedefaultmidpunct}
{\mcitedefaultendpunct}{\mcitedefaultseppunct}\relax
\EndOfBibitem
\bibitem[Mortensen \latin{et~al.}(2014)Mortensen, Raza, Wubs, S{\o}ndergaard,
  and Bozhevolnyi]{mortensen2014generalized}
Mortensen,~N.~A.; Raza,~S.; Wubs,~M.; S{\o}ndergaard,~T.; Bozhevolnyi,~S.~I.
  \emph{Nature Commun.} \textbf{2014}, \emph{5}\relax
\mciteBstWouldAddEndPuncttrue
\mciteSetBstMidEndSepPunct{\mcitedefaultmidpunct}
{\mcitedefaultendpunct}{\mcitedefaultseppunct}\relax
\EndOfBibitem
\bibitem[Cirac{\`\i} \latin{et~al.}(2012)Cirac{\`\i}, Hill, Mock, Urzhumov,
  Fern{\'a}ndez-Dom{\'\i}nguez, Maier, Pendry, Chilkoti, and
  Smith]{ciraci2012probing}
Cirac{\`\i},~C.; Hill,~R.; Mock,~J.; Urzhumov,~Y.;
  Fern{\'a}ndez-Dom{\'\i}nguez,~A.; Maier,~S.; Pendry,~J.; Chilkoti,~A.;
  Smith,~D. \emph{Science} \textbf{2012}, \emph{337}, 1072--1074\relax
\mciteBstWouldAddEndPuncttrue
\mciteSetBstMidEndSepPunct{\mcitedefaultmidpunct}
{\mcitedefaultendpunct}{\mcitedefaultseppunct}\relax
\EndOfBibitem
\bibitem[Yan \latin{et~al.}(2012)Yan, Wubs, and Mortensen]{yan2012effects}
Yan,~W.; Wubs,~M.; Mortensen,~N.~A. Effects of nonlocal response on the density
  of states of hyperbolic metamaterials. SPIE NanoScience+ Engineering. 2012;
  pp 84550V--84550V\relax
\mciteBstWouldAddEndPuncttrue
\mciteSetBstMidEndSepPunct{\mcitedefaultmidpunct}
{\mcitedefaultendpunct}{\mcitedefaultseppunct}\relax
\EndOfBibitem
\bibitem[Wang \latin{et~al.}(2006)Wang, Wu, Lassiter, Nehl, Hafner, Nordlander,
  and Halas]{wang2006symmetry}
Wang,~H.; Wu,~Y.; Lassiter,~B.; Nehl,~C.~L.; Hafner,~J.~H.; Nordlander,~P.;
  Halas,~N.~J. \emph{Proc. Natl. Acad. Sci.} \textbf{2006}, \emph{103},
  10856--10860\relax
\mciteBstWouldAddEndPuncttrue
\mciteSetBstMidEndSepPunct{\mcitedefaultmidpunct}
{\mcitedefaultendpunct}{\mcitedefaultseppunct}\relax
\EndOfBibitem
\bibitem[Mukherjee \latin{et~al.}(2010)Mukherjee, Sobhani, Lassiter, Bardhan,
  Nordlander, and Halas]{mukherjee2010fanoshells}
Mukherjee,~S.; Sobhani,~H.; Lassiter,~J.~B.; Bardhan,~R.; Nordlander,~P.;
  Halas,~N.~J. \emph{Nano Lett.} \textbf{2010}, \emph{10}, 2694--2701\relax
\mciteBstWouldAddEndPuncttrue
\mciteSetBstMidEndSepPunct{\mcitedefaultmidpunct}
{\mcitedefaultendpunct}{\mcitedefaultseppunct}\relax
\EndOfBibitem
\end{mcitethebibliography}

\end{document}